\documentclass[nofootinbib,
  notitlepage,
  aps]
  {revtex4-1}
\usepackage{amsmath,amssymb}
\usepackage{graphicx}
\graphicspath{{figs/}}
\newcommand{\ie}{i.e.,\,}
\newcommand{\eg}{e.g.,\,}
\newcommand{\ud}{\ensuremath{{\rm d}}}

\newcommand{\figref}[1]{figure~\ref{#1}}
\newcommand{\Figref}[1]{Figure~\ref{#1}}
\newcommand{\secref}[1]{section~\ref{#1}}

\newcommand{\appref}[1]{appendix~\ref{#1}}

\newcommand{\abs}[1]{\ensuremath{\left|#1\right|}}
\newcommand{\ave}[1]{\ensuremath{\left\langle #1 \right\rangle}}
\newcommand{\aveE}[1]{\ensuremath{\left\langle #1 \right\rangle_{\rm E}}}
\newcommand{\aveV}[1]{\ensuremath{\left\langle #1 \right\rangle_{\rm V}}}

\newcommand{\meanPhi}{\ensuremath{\bar{\phi}}}
\newcommand{\forceEff}{\ensuremath{F_{\rm eff}}}
\newcommand{\kUV}{\ensuremath{k_{\rm UV}}}
\newcommand{\kIR}{\ensuremath{k_{\rm IR}}}
\newcommand{\mPS}{\ensuremath{m_{\rm PS}}}
\newcommand{\meff}{\ensuremath{m_{\rm eff}}}
\newcommand{\kCut}{\ensuremath{k_{\rm cut}}}
\newcommand{\nCut}{\ensuremath{n_{\rm cut}}}

\newcommand{\mBEC}{\ensuremath{m}}
\newcommand{\lamBEC}{\ensuremath{\lambda}}
\newcommand{\fldBEC}{\ensuremath{\phi_0}}

\newcommand{\vBare}{\ensuremath{V_{\rm bare}}}
\newcommand{\mBare}{\ensuremath{m_{\rm B}}}
\newcommand{\vTree}{\ensuremath{V_{\rm T}}}
\newcommand{\vEff}{\ensuremath{V_{\rm eff}}}

\newcommand{\phiFV}{\ensuremath{\phi_{\rm FV}}}
\newcommand{\phiTV}{\ensuremath{\phi_{\rm TV}}}
\newcommand{\mFV}{\ensuremath{m_{\rm FV}}}
\newcommand{\mTV}{\ensuremath{m_{\rm TV}}}

\newcommand{\fldVar}{\ensuremath{\sigma_{\phi}^2}}
\newcommand{\fldVarPS}{\ensuremath{\sigma_{\rm PS}^2}}
\newcommand{\fldVarBare}{\ensuremath{\sigma^2_{\rm B}}}

\newcommand{\mLoopNL}{\ensuremath{m_{\rm 1-loop,NL}}}
\newcommand{\mLoopIt}{\ensuremath{m_{\rm 1-loop,(1)}}}
\newcommand{\mGaussNL}{\ensuremath{m_{\rm G}}}
\newcommand{\mGaussIt}{\ensuremath{m_{\rm G,(1)}}}

\newcommand{\dk}{\ensuremath{\Delta k}}

\begin{document}

\begin{abstract}
False vacuum decay, a quantum mechanical first-order phase transition in scalar field theories, is an important phenomenon in early universe cosmology. Recently, real-time semi-classical techniques based on ensembles of lattice simulations were applied to the problem of false vacuum decay. In this context, or any other lattice simulation, the effective potential experienced by long-wavelength modes is not the same as the bare potential. To make quantitative predictions using the real-time semi-classical techniques, it is therefore necessary to understand the redefinition of model parameters and the corresponding deformation of the vacuum state, as well as stochastic contributions that require modeling of unresolved subgrid modes. In this work, we focus on the former corrections and compute the expected modification of the true and false vacuum effective mass, which manifests as a modified dispersion relationship for linear fluctuations about the vacuum. We compare these theoretical predictions to numerical simulations and find excellent agreement. Motivated by this, we use the effective masses to fix the shape of a parameterized effective potential, and explore the modeling uncertainty associated with  non-linear corrections. We compute the decay rates in both the Euclidean and real-time formalisms, finding qualitative agreement in the dependence on the UV cutoff. These calculations further demonstrate that a quantitative understanding of the rates requires additional corrections.  
\end{abstract}
  
\title{Mass Renormalization in Lattice Simulations of False Vacuum Decay}
\author{Jonathan Braden}
\affiliation{Canadian Institute for Theoretical Astrophysics, University of Toronto, 60 St.\ George Street, Toronto, ON, M5S 3H8, Canada}
\author{Matthew C.\ Johnson}
\affiliation{Department of Physics and Astronomy, York University, Toronto, ON, M3J 1P3, Canada}
\affiliation{Perimeter Institute for Theoretical Physics, 31 Caroline St.\ N, Waterloo, ON, N2L 2Y5, Canada}
\author{Hiranya V.\ Peiris}
\affiliation{Department of Physics and Astronomy, University College London, Gower Street, London, WC1E 6BT, UK}
\affiliation{The Oskar Klein Centre for Cosmoparticle Physics, Department of Physics, Stockholm University, AlbaNova, Stockholm, SE-106 91, Sweden}
\author{Andrew Pontzen}
\affiliation{Department of Physics and Astronomy, University College London, Gower Street, London, WC1E 6BT, UK}
\author{Silke Weinfurtner}
\affiliation{School of Mathematical Sciences, University of Nottingham, University Park, Nottingham, NG7 2RD, UK}
\affiliation{Centre for the Mathematics and Theoretical Physics of Quantum Non-Equilibrium Systems, University of Nottingham, Nottingham, NG7 2RD, UK}

\maketitle

\section{Introduction}\label{sec:introduction}
False vacuum decay is a fascinating phenomenon in quantum field theory involving many of its  most challenging aspects: it is  dynamical, nonlinear, and non-perturbative.
It has applications to a variety of fundamental scenarios in cosmology, and is also relevant to a broad range of condensed matter systems.
While direct experimentation with cosmological systems is beyond current technologies, recent proposals to create analogue relativistic false vacuum systems in the lab~\cite{Fialko:2014xba,Fialko:2016ggg,Braden:2017add,Billam:2018pvp,Braden:2019vsw,Billam:2020xna,Ng:2020pxk,Billam:2021qwt,Billam:2021nbc,Billam:2021psh} provide the exciting possibility of experimentally probing some aspects of false vacuum decay.
To make maximal use of such experiments and compare to theory, a thorough understanding of the theoretical modeling of false vacuum decay is required.
The calculations in this work demonstrate the importance of understanding renormalization effects on the properties of the false vacuum.

Relativistic vacuum decay is the process by which a large region of space trapped in a local potential minimum (\ie a false vacuum) transitions to a new state that is no longer localized in the potential minimum.
The process is expected to occur via the formation of bubbles within the ambient false vacuum.
Inside these `true vacuum' bubbles, the field is on the opposite side of the barrier and no longer trapped in the false vacuum. 
The bubbles then expand, with the speed of the walls rapidly approaching the speed of light.
In many contexts the bubbles eventually coalesce, resulting in a phase of bubble collisions and the conversion of the initial false vacuum into an (excited) state localized around the true vacuum.
In other cases, such as when the space in the false vacuum  regions is undergoing rapid accelerated expansion, the bubbles fail to coalesce, leading to an incomplete phase transition and a state consisting of the ambient false vacuum with bubbles embedded within.

The standard approach to vacuum decay, referred to as the bounce formalism~\cite{Coleman:1977py,Callan:1977pt}, works in Euclidean time, thus obscuring the fundamentally time-dependent nature of the problem.  In this formalism, a high degree of symmetry is imposed on the nucleating bubbles, with contributions from inhomogeneities assumed perturbative in $\hbar$ and relegated to (frequently neglected) correction terms.  
Furthermore, bubble nucleations are interpreted as quantum tunneling events.  Therefore, there is no analogue of classical time evolution to describe the formation of a bubble.  The basic input in this approach is to specify a potential for the mean field, which we will call the tree-level Euclidean potential and denote $\vTree$.  From this we can compute expected rates of bubble nucleation and the typical profile of a nucleated bubble.

Ref~\cite{Braden:2018tky} introduced an alternative approach to the problem of vacuum decay, which instead works in real-time by evolving realizations of the false vacuum forward in time (see also~\cite{Hertzberg:2019wgx,Hertzberg:2020tqa,Blanco-Pillado:2019xny,Mou:2019gyl,Ai:2019fri,Michel:2019nwa,Huang:2020bzb,Pirvu:2021roq} for subsequent related work).
Analogous techniques are used in a variety of other contexts, such as: preheating at the end-of-inflation~\cite{Felder:2000hq,Micha:2004bv,Frolov:2008hy,Easther:2010qz}, evolution of dilute gas cold atom Bose-Einstein condensates~\cite{PhysRevA.58.4824,PhysRevA.68.053604}, relativistic heavy ion collisions~\cite{Berges:2015ixa,Berges:2020fwq}, and nonequilibrium and thermalization properties of relativistic fields~\cite{Borsanyi:2000ua, Borsanyi:2001rc, Borsanyi:2002tm}.
In this approach, the problem is treated with classical statistical field theory by sampling and evolving initial configurations of the false vacuum.
In this framework, the purely classical time-evolution leads to the dynamical formation of ``true vacuum'' bubbles from the (Gaussian) realizations of the false vacuum.  In analogy to the Euclidean view, we will refer to the emergence of a bubble in these simulations as the decay of the false vacuum.  These bubbles then expand and eventually coalesce, causing the system to transition from the false vacuum to a state localized around the true vacuum.  The distribution of bubble nucleation times are well-modelled by an exponential decay law, allowing for a straightforward determination of the decay rate.  Thus, these simulations give an explicitly time-dependent description of the formation of bubbles, providing a classical time-dependent connection between the initial Gaussian vacuum state and the subsequent phase of an expanding bubble.  In particular, we do not assume an initial bubble shape (unlike previous investigations~\cite{PhysRevD.60.125003}), nor do we assume that a ``tunneling'' event describes the sudden appearance of a bubble.
Further, since we have access to much more fine-grained information, many additional questions that are inaccessible in the Euclidean framework can be explored, such as the correlation between bubble nucleation events~\cite{Pirvu:2021roq}.
This approach fully captures nonlinearity (\ie backreaction and rescattering effects) and places no symmetry assumptions on the resulting solutions, but does not incorporate interference effects between different initial realizations. 
In addition to a potential, which we will call the bare lattice potential and denote $\vBare$, the real-time approach requires an assumption about the statistics of the initial vacuum field fluctuations. 

\begin{figure}[t!]
  \includegraphics[width=3.5in]{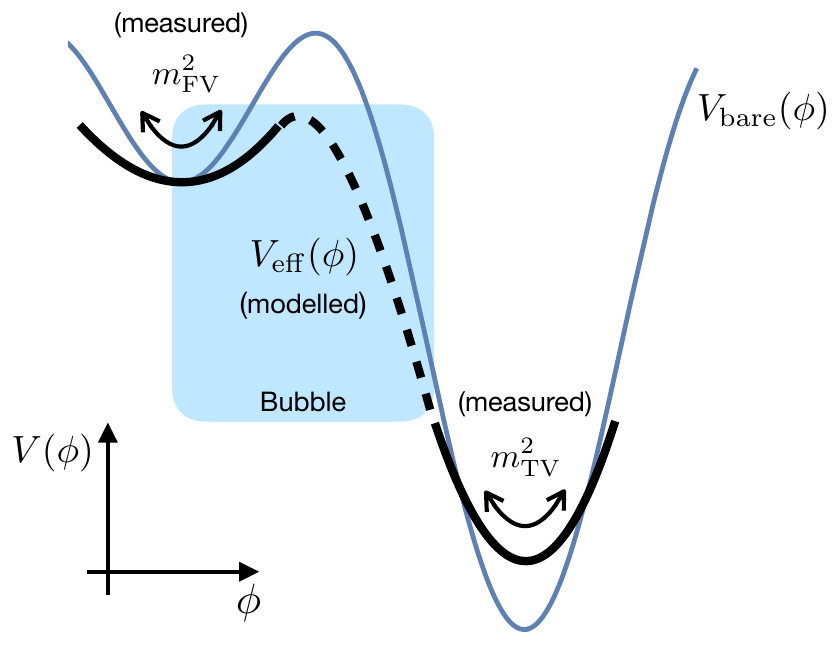}
  \caption{An illustration of how we determine the effective potential $\vEff$ seen by long-wavelength fluctuations. We develop analytic and empirical techniques to measure the corrections to the curvature about the False ($\mFV^2$) and True ($\mTV^2$) vacua defined by the bare potential $\vBare$ (light solid blue line). Modeling $\vEff$ by an expansion that respects the periodic symmetry of the bare potential, we use the measured $\mFV^2$ and $\mTV^2$ to fix two free parameters. We can then explore the modeling uncertainty arising from additional terms in the expansion (relevant in the shaded region of the potential), which affect the properties of the critical bubble (instanton solution) that determines the Euclidean rate obtained from $\vEff$.}
  \label{fig:potential-cartoon}
\end{figure}

Summarizing, in the Euclidean approach the bubbles are assumed to be highly symmetric and in the standard interpretation appear via a quantum tunneling process with no classical real-time description.
Meanwhile, the semi-classical real-time approach makes no symmetry assumptions about the bubbles and provides a classical description of the moment of nucleation.  However, the latter approach is unable capture interference between different histories.
Given the drastically different interpretations and approximations used in the two approaches, we would like to understand if they are different descriptions of the same decay process or if they describe different mechanisms by which bubbles nucleate in the false vacuum.
A first step is to compare the decay rates between the two approaches.
If we identify the tree-level Euclidean potential with the bare lattice potential, we find that the decay rate extracted from the real-time approach is much larger than that of the Euclidean approach.
Specifically, the log of the decay rate decreases more slowly as the amplitude of the quantum fluctuations is decreased (relative to the width of the barrier) for the real-time simulations than the Euclidean prediction.
This raises an interesting puzzle about the relationship between the two approaches, pointed out in Refs.~\cite{Braden:2018tky,Hertzberg:2020tqa}.

However, as noted in Ref.~\cite{Braden:2018tky}, identifying the tree-level Euclidean potential with the bare lattice potential is not correct because it ignores important renormalization effects associated with the lattice simulations. To make a proper comparison of the decay rates, we must determine which tree-level Euclidean potential corresponds to the bare lattice potential employed by the real-time formalism. The main focus of this paper is to make progress towards this goal.

In more detail, while the bare lattice potential correctly encodes the dynamics for the full inhomogeneous field realizations, to obtain the dynamics of the mean field alone, we must integrate out the inhomogeneities.  This results in a modification of the mean field dynamics, similar to the effects of renormalization in quantum field theory.
Since the Euclidean potential encodes the dynamics of the mean field, it is reasonable to conjecture that the corrections to the bare lattice potential should be included in the mapping between the lattice and Euclidean calculations.
In what follows, we will refer to this as the effective lattice potential and denote it $\vEff$.

We determine the effective force felt by the spatially-averaged mean field assuming Gaussianity of the fluctuations, an improvement on a one-loop approximation.
From this, we compute the effective mass squared (\ie effective potential curvature) for mean field values located at symmetric minima of the bare potential. 
Unfortunately, the situation is significantly more complicated when the potential is no longer symmetric about the minimum, or if we displace the mean field from the potential minimum.
To explore the potential impacts of these deformations on the Euclidean decay rate, we write down some simple parametrizations of the full effective potential, with tunable parameters that can be calibrated using either our analytic calculations of the effective mass, or measurements made in simulations. An illustration of this  approach can be found in~\figref{fig:potential-cartoon}.
While there is some uncertainty in our theoretical modeling, we place reasonable bounds on the sensitivity of the Euclidean decay rate to the details of the short-wavelength fluctuations.

Although we focus on false vacuum decay, the formalism outlined here is universally applicable to tackling renormalization issues in real-time semiclassical stochastic lattice simulations.
Therefore, the methods developed here may also be of use in other settings, such as generating initial conditions for preheating simulations.

The remainder of this paper is organised as follows. In~\secref{sec:framework} we review the real-time lattice framework and the particular potential we consider in this paper.  In~\secref{sec:mass-renorm} we compute the expected renormalization of the field mass and compare it to results from lattice simulations, finding excellent agreement across a range of parameters.  We then explore the possible impacts of these deformations on the predicted Euclidean decay rate in~\secref{sec:euclidean-actions}.  Finally, we describe how some simple modifications of the initial fluctuations modify the real-time decay rates extracted from lattice simulations.  We then conclude in~\secref{sec:conclusion}.  Details of our effective potential modeling to compute the Euclidean decay rates are provided in~\appref{app:potential-modelling}.
Unless otherwise stated, we work in units with $\hbar =  c = 1$ throughout.

\section{Stochastic Lattice Simulations}\label{sec:framework}
We work within the semiclassical stochastic lattice framework, which we now briefly review (see Ref.~\cite{Braden:2018tky} for further details).
The key assumption is to model quantum effects by sampling field and momentum configurations of the quantum vacuum.
Generating samples of the fully self-consistent interacting vacuum is a highly nontrivial undertaking.
Therefore, we approximate the vacuum state as Gaussian and initialize
\begin{subequations}\label{eqn:fluc-expansion}
\begin{align}
  \phi(x,t=0)       &= \phi_{\rm init} + \delta\hat{\phi} = \phi_{\rm init} + \frac{1}{\sqrt{V}}\sum_{n\neq 0}^{\abs{n}\leq \nCut} \frac{\hat{\alpha}_n}{\sqrt{2\omega_n}}e^{i\frac{2\pi}{L}n x} \\
  \Pi(x,t=0)        &= \delta\hat{\Pi} = \frac{1}{\sqrt{V}}\sum_{n\neq 0}^{\abs{n}\leq\nCut} \sqrt{\frac{\omega_n}{2}}\hat{\beta}_ne^{i\frac{2\pi}{L}n x} \, ,
\end{align} 
\end{subequations}
where $\hat{\alpha}_n$ and $\hat{\beta}_n$ are independent complex Gaussian random deviates satisfying $\aveE{\abs{\alpha_n}^2} = 1 = \aveE{\abs{\beta_n}^2}$, $\alpha_{-n}=\alpha_n^*$, $\beta_{-n}=\beta_n^*$, and $\aveE{\cdot}$ denotes an ensemble average. $V$ is the simulation volume and $\omega_n$ is the oscillation frequency of the $n$th mode.  
Provided the fluctuations primarily probe the quadratic region of the local potential minimum, we expect this to be a good approximation to the true vacuum state.
Since we work on a discrete lattice of finite size, we have truncated the spectrum above some wavenumber $\kCut \equiv \frac{2\pi}{L}\nCut$ below the Nyquist wavenumber $k_{\rm Nyq} = \frac{\pi}{dx}$ where $dx$ is the grid spacing.
For the results presented here, we used convergence testing to verify that $\kCut$ was sufficiently large to capture all relevant dynamical evolution of the fluctuations.
In particular, for our choice of lattice and model parameters this demonstrates that there is minimal transfer of power between the ``short-wavelength'' modes with $k>\kCut$ and ``long-wavelength'' modes with $k<\kCut$.  Note that since our lattice has a finite volume, initializing modes directly in Fourier space does not exactly reproduce the real space two-point correlation function.  Therefore, we also confirmed the finite box size has a negligible effect on our results, which is expected since the mass of the field cuts off the infrared fluctuation power per $\ln k$ bin.\footnote{To empirically check this, we compared our results using lattices of different spatial extents but fixed spatial resolution.}

We then propagate these initial conditions forward in time using the classical equations of motion
\begin{subequations}
\begin{align}
  \frac{\ud\phi}{\ud t} &= \Pi \\
  \frac{\ud\Pi}{\ud t}  &= \nabla^2\phi - \frac{\partial V}{\partial\phi} \, .
\end{align}
\end{subequations}
As in our previous work~\cite{Braden:2018tky}, we use a 10th order Gauss-Legendre integrator for time-evolution, and a Fourier collocation discretization of space (see the appendices of Ref.~\cite{Bond:2015zfa} for more details).
Expectation values of quantum operators are approximated by a classical ensemble average over realizations of the initial fluctuations.

The philosophy of this approach is to model quantum effects through the randomly drawn initial conditions $\delta\hat{\phi}$ and $\delta\hat{\Pi}$, but not directly in the time evolution.  Heuristically, we are enforcing the uncertainty principle in the initial conditions by drawing individual field and momentum Fourier modes as Gaussian random deviates with the appropriate width.  Meanwhile, we are neglecting interference effects in the path integral between different classically-evolved trajectories, which are not captured by our classical time evolution.  This approach is expected to work well when the ``quantum'' fluctuations are initially linear, and at the level of reproducing operator expectation values can be shown to incorporate the first two orders of a systematic expansion of the full quantum dynamics in $\hbar$ (see for example the supplemental material of Ref.~\cite{Braden:2018tky}).
In the above, we have restricted ourselves to spatially homogeneous background solutions that are at rest.
Via appropriate generalizations of the mode functions, this can be extended to the case with non-zero initial field momentum, or spatial inhomogeneity.
Although we will not pursue it here, modifying the covariance structure of the $\hat{\alpha}_k$'s and $\hat{\beta}_k$'s allows us to explore nonvacuum states, such as squeezed states or thermal states.

The stochasticity, therefore, refers to the particular realization of the initial conditions, not additional noise terms appearing during the time-evolution.
This approach is related, at least in spirit, to alternative stochastic approaches such as: stochastic tunneling (see \eg Ref.~\cite{Linde:1991sk}), stochastic inflation (see \eg Refs.~\cite{Starobinsky:1982ee,Starobinsky:1986fx,Salopek:1990jq}) or the Langevin approach to thermal field theory on the lattice (see \eg~\cite{Gleiser:1993ea,Borrill:1994nk,Borrill:1996uq}), where the short-wavelength modes are modeled as additional noise contributions during the time evolution.
The primary difference is that rather than externally imposing a model for these additional noise terms, we are self-consistently solving for their dynamics as the system evolves.
For example, suppose we restrict ourselves to consideration of the spatial average of the field, which we will refer to as the zero mode.
From the viewpoint of the effective dynamics of the zero mode, all of the realized inhomogeneous modes effectively act as stochastic contributions to the dynamics. 
However, rather than having to explicitly model the dynamics of these bath modes, we are self-consistently evaluating the statistical properties of the noise by evolving the realized modes.

In this paper we focus on the analogue false vacuum decay potential
\begin{equation}\label{eqn:bec-potential}
  \vBare = \mBEC^2\fldBEC^2\left[\cos\left(\frac{\phi}{\fldBEC}\right) - 1 + \frac{\lamBEC^2}{2}\sin^2\left(\frac{\phi}{\fldBEC}\right)\right]
\end{equation}
with corresponding potential force
\begin{equation}\label{eqn:bec-force}
  \vBare' = \mBEC^2\fldBEC \left[-\sin\left(\frac{\phi}{\fldBEC}\right) + \frac{\lamBEC^2}{2}\sin\left(2\frac{\phi}{\fldBEC}\right)\right] \, .
\end{equation}
For $\lamBEC^2 > 1$, this potential possesses an infinite sequence of false vacuum minima at $\phiFV = 2\pi n \fldBEC$ and a corresponding sequence of true vacuum minima at $\phiTV = (2n+1)\pi \fldBEC$ with $n\in \mathbb{Z}$.
Since we have in mind the application of this potential to false vacuum decay, in this paper we will primarily concentrate on states initially localized around the false vacuum $\phi = 0$.
To make the most direct connection with existing work~\cite{Braden:2018tky,Hertzberg:2020tqa,Pirvu:2021roq}, we will also limit ourselves to one spatial dimension.
We will provide a more detailed exploration in higher dimensions in a future publication.

\section{Renormalization of the false vacuum}\label{sec:mass-renorm}
As outlined above, we are interested in developing an effective potential description of our lattice simulations,
with the goal of connecting to the potential appearing in Euclidean calculations of the false vacuum decay rate.
The appropriate dynamical variable is therefore the mean field, $\meanPhi \equiv \ave{\phi}$.
For lattice simulations, rather than work directly with the potential, it is more convenient to study the additional force experienced by $\meanPhi$.  Roughly, this corresponds to studying the derivative of the effective potential.

To obtain the effective dynamics of the mean field, we average the equation of motion
\begin{equation}
  \ddot{\meanPhi} + \ave{\vBare'} = 0 \, ,
\end{equation}
where we have assumed that the mean field is spatially homogeneous so we can discard the Laplacian term.  For now we allow $\ave{\cdot}$ to represent either an ensemble $\aveE{\cdot}$ or a volume $\aveV{\cdot}$ average.  In a given realization (\ie simulation) these two quantities deviate from each other, but if we average the volume average over an ensemble of (finite-volume) numerical simulations, we expect it to converge to the full ensemble average.  For notational convenience, we define the (negative of the) effective force acting on the mean field
\begin{equation}
  \forceEff \equiv \ave{\vBare'} \, .
\end{equation}
If $\forceEff$ is a function of the mean field alone we can identify $V'_{\rm eff} = \forceEff$ and integrate to obtain an effective potential, although this need not hold in general situations.

We now define $\phi(x,t) = \meanPhi(t) + \delta\phi(x,t)$ and assume that we can Taylor expand the potential around the instantaneous mean field value $\meanPhi$, transforming $\ave{\vBare'}$ into a weighted sum over the moments of $\delta\phi$.
One can re-sum the Gaussian contributions, and for the analogue BEC potential~\eqref{eqn:bec-potential} we obtain
\begin{equation}\label{eqn:force-gaussian-resum}
  \frac{\forceEff(\meanPhi)}{\mBEC^2\fldBEC} = e^{-\frac{\fldVar}{2\fldBEC^2}}\left[-\sin\left(\frac{\meanPhi}{\fldBEC}\right) + e^{-\frac{3}{2}\frac{\fldVar}{\fldBEC^2}}\frac{\lamBEC^2}{2}\sin\left(2\frac{\meanPhi}{\fldBEC}\right)\right] + \frac{F_{\rm NG}}{\mBEC^2\fldBEC} \, ,
\end{equation}
where $\fldVar \equiv \ave{\delta\phi^2}$ and $F_{\rm NG}$ encodes all non-Gaussian contributions to the one-point distribution of $\delta\phi$.
In the following theoretical development, we will neglect the effects of these non-Gaussian corrections.
Under the assumptions of Gaussianity and a Taylor-expandable bare potential, we have therefore reduced the expression for the effective force acting on the mean field to the theory of a single function of the mean field, $\fldVar(\meanPhi)$.

\subsection{Vacuum Fluctuations}
To make further progress we must provide a model for $\fldVar$.
Since we are interested in relativistic vacuum fluctuations, we assume
\begin{subequations}
\begin{align}
  \aveE{\delta\tilde{\phi}_k\delta\tilde{\phi}_{k'}^*} &= \frac{\dk}{2\pi}\frac{1}{2\sqrt{k^2+\mPS^2}}\delta_{k,k'} \\
  \aveE{\delta\tilde{\Pi}_k\delta\tilde{\Pi}_{k'}^*} &= \frac{\dk}{2\pi}\frac{\sqrt{k^2+\mPS^2}}{2}\delta_{k,k'} \\
  \aveE{\delta\tilde{\phi}_k\delta\tilde{\Pi}_{k'}^*} &= 0\, .
\end{align}
\end{subequations}
We work in the discrete limit where $L$ is the side length of our simulation volume and $\dk = \frac{2\pi}{L}$.
In linear perturbation theory we would identify $\mPS^2 = \vBare''$.
However, the presence of the realized fluctuations leads to a deformation of the vacuum state.
A key result of this work is to calculate an improved value of $\mPS^2$ describing the corrected vacuum.
More generally, we can allow $\mPS^2$ to be a free parameter describing the initial state of our system, and then explore the subsequent response of the system to this class of (generally nonvacuum) initial conditions.
In the remainder of the paper, we will refer to this class of initial conditions as \emph{pseudovacuum} initial conditions.

It is straightforward to obtain the field variance in these pseudovacuum states, which we denote $\fldVarPS$.  Restricting to one spatial dimension, we have
\begin{equation}\label{eqn:variance-vac-1d}
  \fldVarPS(\mPS^2) = \frac{1}{2\pi}\ln\left(\frac{\kUV + \sqrt{\kUV^2+\mPS^2}}{\kIR+\sqrt{\kIR^2+\mPS^2}}\right) \, ,
\end{equation}
where we have assumed UV ($\kUV$) and IR ($\kIR$) cutoffs on the spectrum.
In the context of lattice simulations, the UV cutoff arises naturally from the spectral truncation~\eqref{eqn:fluc-expansion} required to initialize our fluctuations.
Meanwhile, the IR cutoff can arise either from the existence of the fundamental mode in our simulations or the dynamical emergence of localized spatial structures.
Assuming the fundamental mode is the relevant IR scale, and accounting for the implicit centered binning of our Fourier modes, we have
\begin{equation}\label{eqn:spectral-limits}
   \kIR = \frac{\pi}{L} \qquad {\rm and} \qquad \kUV = \frac{2\pi}{L}\left(n_{\rm cut} + \frac{1}{2}\right) = \kCut + \frac{\pi}{L} \, ,
\end{equation}
which will be confirmed in~\secref{subsec:mass-measurements} below.
We are typically interested in the limits $\kUV^2 \gg \abs{\mPS^2}$ and $\kIR^2 \ll \abs{\mPS^2}$, for which
\begin{equation}
  \fldVarPS \approx \frac{1}{4\pi}\ln\left(\frac{4\kUV^2}{\mPS^2}\right) \, .
\end{equation}
Within this framework, we have swapped the unknown variance $\fldVarPS$ for the unknown effective mass $\mPS^2$.

\subsection{Renormalized False Vacuum Mass}
Now that we have an understanding of how the fluctuations modify the dynamics of the mean field, we want to determine how the effective mass of the field is deformed from its bare value. 
There are two masses appearing in our formulation: the mass $\mPS$ that determines the variance of the fluctuations; and a dynamical mass associated with the curvature of the effective potential, which we will denote $\meff$.
Assuming the system is in a pseudovacuum state, we have
\begin{align}\label{eqn:m2eff-equation}
  \frac{\meff^2}{\mBEC^2} &\equiv \frac{\ud\forceEff}{\ud\meanPhi}(\phiFV;\fldVarPS(\mPS^2)) = \lamBEC^2e^{-\frac{2\fldVarPS}{\fldBEC^2}} - e^{-\frac{\fldVarPS}{2\fldBEC^2}} \, ,
\end{align}
where $\fldVarPS$ is given by~\eqref{eqn:variance-vac-1d}.
The emergence of a nonlinear algebraic equation~\eqref{eqn:m2eff-equation} for $m^2_{\rm eff}$ is a consequence of the evenness of the potential about the minimum, which leads to two simplifications:
(i) the location of the false vacuum does not shift in response to the presence of the fluctuations, and (ii) derivatives of $\sigma^2_{\rm PS}$ with respect to $\bar{\phi}$ do not appear.

In the self-consistent vacuum state, the potential curvature ($\meff^2$) must match the squared effective mass in the initial conditions ($\mPS^2$).
Denoting this special value by $\mGaussNL^2$, we must solve
\begin{equation}\label{eqn:m2eff-gauss-nl}
  \frac{\mGaussNL^2}{\mBEC^2} = \lamBEC^2e^{-\frac{2\fldVar(\mGaussNL^2)}{\fldBEC^2}} - e^{-\frac{\fldVar(\mGaussNL^2)}{2\fldBEC^2}} \, .
\end{equation}
In the remainder of the paper, we will call this the nonlinear Gaussian approximation.

In the absence of fluctuations, we have $\meff^2 = \vBare''(\phiFV) = \mBEC^2\left(\lamBEC^2-1\right) \equiv \mBare^2$.
We obtain a simple analytic approximation by substituting $\fldVarPS\left(\mBare^2\right) \equiv \fldVarBare$ into the RHS of~\eqref{eqn:m2eff-gauss-nl}
\begin{equation}\label{eqn:m2eff-first-iter}
  \frac{\mGaussIt^2}{m^2} = \lambda^2e^{-2\frac{\fldVarBare}{\fldBEC^2}} - e^{-\frac{1}{2}\frac{\fldVarBare}{\fldBEC^2}} \, .
\end{equation}
We refer to this as the Gaussian bare mass approximation.

We can also obtain a perturbative approximation in the fluctuation amplitudes by first expanding~\eqref{eqn:m2eff-equation} in $\fldVarPS$. To leading nontrivial order we have
\begin{equation}\label{eqn:m2eff-equation-one-loop}
  \frac{\meff^2}{\mBEC^2} \approx \mBare^2 + \frac{V''''(\phiFV)}{2}\fldVarPS = \lamBEC^2 - 1 - \frac{(4\lamBEC^2-1)}{2}\frac{\fldVarPS}{\fldBEC^2} \, ,
\end{equation}
which we refer to as the one-loop approximation.
Equating $\meff^2$ and $\mPS^2$ as before, we can then solve the resulting nonlinear equation.
We will call this the nonlinear one-loop approximation and denote $\mLoopNL^2$.
Alternatively, we can further approximate $\fldVarPS \approx \fldVarBare$ to obtain
\begin{equation}\label{eqn:m2eff-one-loop-it}
  \frac{\mLoopIt^2}{\mBEC^2} \approx \lamBEC^2 -1 - \left(\frac{4\lamBEC^2-1}{4\pi\fldBEC^2}\right)\ln\left(\frac{\kUV+\sqrt{\kUV^2+\lamBEC^2-1}}{\kIR + \sqrt{\kIR^2 + \lamBEC^2-1}}\right) \, ,
\end{equation}
which we will refer to as the one-loop bare mass approximation.

\begin{figure}
  \includegraphics[width=7.05in]{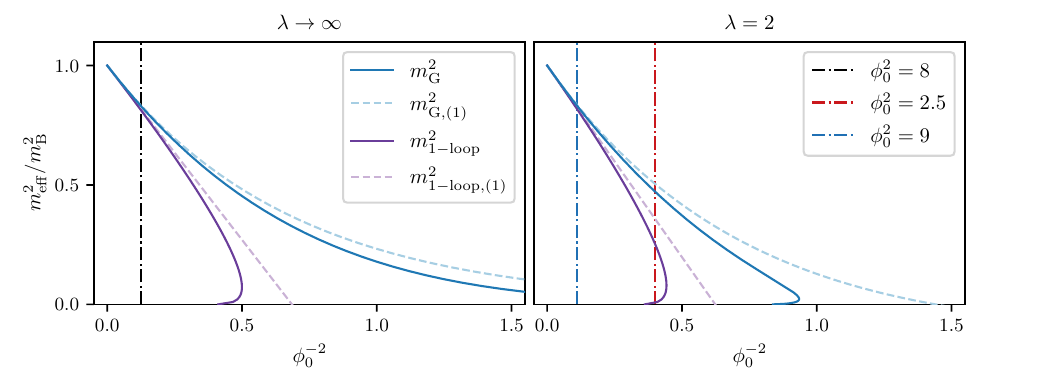}
  \caption{Analytic predictions for the effective mass of the false vacuum minimum, including the effect of the realized vacuum fluctuations. In the left plot we take $\lamBEC \to \infty$, and in the right plot $\lamBEC=2$.   Here we have included the full nonlinear solution in the Gaussian approximation (\emph{dark solid blue line}) and one-loop approximation (\emph{dashed light blue line}).  We also include the bare mass approximations for the Gaussian (\emph{solid dark purple line}) and one-loop (\emph{dashed light purple line}) approximations.  To compare directly with our lattice simulations below, we have taken $\frac{\kIR}{\lamBEC\mBEC} = \frac{\pi}{128}$ and $\frac{\kUV}{\lamBEC\mBEC} = \frac{2\pi}{128}\times 1023.5$ in the left panel; and $\frac{\kIR}{\mBEC} = \frac{\pi}{64}$ and $\frac{\kUV}{\mBEC} = \frac{2\pi}{64}\times 511.5$ in the right panel.  While all approximation schemes agree in the small fluctuation amplitude limit, $\fldBEC^{-2} \ll 1$, they begin to deviate significantly once $\fldVar$ becomes of order $\fldBEC^2$ and additional terms in the expansion are required.  The vertical dot-dashed lines indicate the particular choices of $\fldBEC$ used in~\figref{fig:variance-plot} (\emph{left panel}) and~\figref{fig:dispersion-plot} (\emph{right panel}) below.}
  \label{fig:m2eff-prediction}
\end{figure}

These predictions are illustrated in~\figref{fig:m2eff-prediction} as $\fldBEC^{-2}$ is varied, while holding fixed the bare potential parameter $\lamBEC^2$, as well as the UV and IR spectral cutoffs.
Since $\fldVar/\fldBEC^2$ controls the amount of nonlinear potential structure probed by the fluctuations, as $\fldBEC \to \infty$ we return to the limit of the bare lattice potential.  For $(\fldVar/\fldBEC^2) \ll 1$ all of the approximation schemes closely match each other.
However, as $\fldBEC$ is decreased and the fluctuations probe more of the nonlinear potential structure, the curves begin to deviate significantly.

For the potential considered here, we find that the one-loop approximation always overestimates the magnitude of the mass correction.  Further, we find the one-loop bare mass approximation (where the field variance is estimated as that in the bare vacuum) is more accurate than a full nonlinear solution to the one-loop equation~\eqref{eqn:m2eff-equation-one-loop}.
Presumably, this is because the nonlinear one-loop solution is implicitly including higher orders in $\fldVar$ in the solution, while not self-consistently including these higher orders in the defining equation.
Comparing to the nonlinear Gaussian approximation, we see the Gaussian bare mass approximation~\eqref{eqn:m2eff-first-iter} provides a significant improvement over either of the one-loop approximations.

\subsection{Lattice Measurements of the Renormalized Mass}\label{subsec:mass-measurements}
We now test the preceding analytic predictions directly in lattice simulations by studying the time-evolution of field realizations from pseudovacuum initial conditions.
To avoid causal interaction of a point with itself, we integrate for a single light-crossing time (\ie set our integration time to the box length $L$).
Since we want to study properties of the false vacuum, in any given realization we must restrict ourselves to times before a bubble nucleates.
Therefore, in this section we take $\lambda^2 \gg 1$ to ensure that most of our simulations do not decay (\ie nucleate a bubble) within a light crossing time.
The rare simulations that decayed were excluded from the ensemble averages.
We chose lattice cutoffs $\kCut$ sufficiently large to ensure that the flow of fluctuation power into UV modes beyond $\kCut$ was strongly suppressed for the duration of the simulations.
As a complementary check, we confirmed that our results were insensitive to changes in the grid spacing at fixed side length $L$ and spectral cutoff $\kCut$.

\begin{figure}
  \includegraphics[width=7.05in]{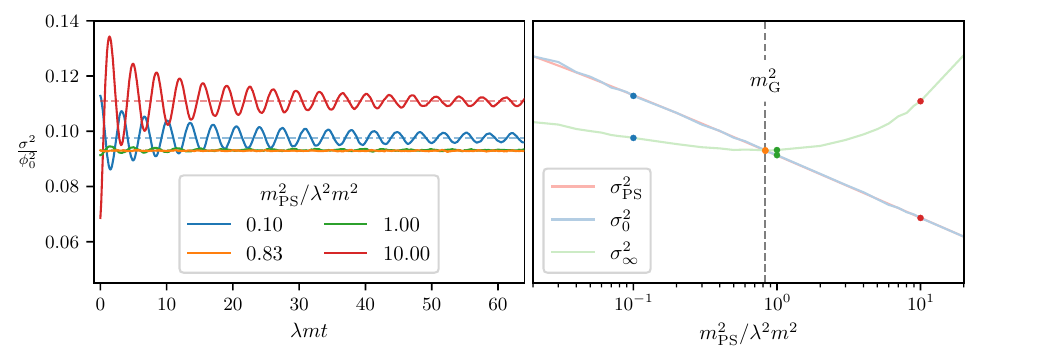}
  \caption{Temporal properties of the field variance $\fldVar \equiv \ave{\delta\phi^2}$ as we vary the mass $m^2_{\rm PS}$ in our initial pseudovacuum states.  To prevent bubble nucleations, we consider the case $\lamBEC \to \infty$, where the analogue BEC model reduces to the sine-Gordon model.  For illustration, we have taken $\fldBEC = 2\sqrt{2}$, resulting in a predicted effective mass $m^2_{\rm G} \approx 0.83\lamBEC^2\mBEC^2$. \emph{Left}: Time-evolution of $\fldVar$ for a few representative choices of initial states parameterized by $\mPS^2$.  The blue line shows a small $\mPS^2$ relative to the bare mass, the orange line the corrected $\mPS^2 = \mGaussNL^2$, the green line the bare $\mPS^2 = \mBare^2$, and the red line a large $\mPS^2$ relative to the bare mass. We see that $\fldVar$ undergoes damped oscillations around an asymptotic value $\sigma^2_{\infty}$ that typically differs from the variance of the pseudovacuum initial state; only in the case of $\mPS^2 = \mGaussNL^2$ is the asymptotic value $\sigma^2_{\infty}$ approximately equal to $\sigma^2_0$.  The numerically estimated asymptotic values are indicated by horizontal dashed lines. \emph{Right}: A comparison between the initial variance $\sigma^2_{0}$ (\emph{light blue line}) and late-time asymptotic variance $\sigma^2_{\infty}$ (\emph{light green line}) as the mass of the initial pseudovacuum fluctuations is varied.  For reference, our Gaussian resummed prediction $\mGaussNL^2$ is shown as a vertical dashed line.  Results for the specific $\mPS^2$ values show in the left plot are indicated with colored dots.  We see the Gaussian prediction for the vacuum corresponds to the special point $\sigma^2_0 = \sigma^2_{\infty}$, which also coincides with the minimum of the asymptotic variance.  Our analytic result for the pseudovacuum variance~\eqref{eqn:variance-vac-1d} is shown as a \emph{light red} line.  We used numerical ensembles of $1000$ lattice simulations in a box of side length $\lamBEC\mBEC L = 128$, $N=2048$ lattice sites, and spectral cutoff $\nCut = 1023$.  We verified the results are insenstive to the choice of grid spacing at fixed $L$ and $\nCut$, and that the dependence on the box size was that expected from adjusting $\kIR$.  We empirically estimated the sampling errors and found they we comparable to the width of the lines, consistent with the agreement between the $\sigma_0^2$ (empirical) and $\fldVarPS$ (theory) lines.}
  \label{fig:variance-plot}
\end{figure}

Given the preceding predictions for the shift in effective mass of the field, we expect a corresponding shift in the effective oscillation frequency of small amplitude modes.
Since the oscillation frequency sets the fluctuation spectrum for linear vacuum fluctuations, a self-consistent determination of the vacuum fluctuations must account for this effect.
In~\figref{fig:variance-plot} we show a coarse-grained demonstration of this by studying the variance of the field fluctuations in the simulations as we allow $\mPS^2$ in our pseudovacuum initial conditions to vary.
If the fluctuations begin in the vacuum, they should have time-stationary statistics.\footnote{For the false vacuum, we mean that the fluctuation statistics remain time-translation invariant on time-scales much shorter than the decay time of the metastable state.}
In particular, $\fldVar$ should be time-independent, at least on time scales much shorter than the decay time of the metastable state.
The left panel of~\figref{fig:variance-plot} shows the time-evolution of $\fldVar$ for a few select values of $\mPS^2$.
We see that in all cases considered here, $\fldVar$ initially experiences a rapid transient before undergoing damped oscillations around a shifted time-averaged mean $\sigma^2_{\infty}$.
This suggests that the initial excited state has settled into some sort of statistically stationary state that differs from the initial pseudovacuum.
Crucially, when we choose $\mPS^2$ to be consistent with our prediction for the modified vacuum mass, the field variance is found to be in a stationary state as expected, and does not undergo damped oscillations.
This is further illustrated in the right panel of the figure, where we show the initial field variance $\sigma^2_0$ and late-time field variance $\sigma^2_{\infty}$ as a function of the initial $\mPS^2$.
For reference, we also include the analytic result for the initial variance $\fldVarPS$ in~\eqref{eqn:variance-vac-1d} with cutoffs $\kIR$ and $\kUV$ given by~\eqref{eqn:spectral-limits}.
The excellent agreement with the empirically-measured curve confirms the relationship between the lattice parameters and cutoffs in the continuum calculations.
A necessary condition for the initial state to be a vacuum is that it be time-stationary.
Therefore, the initial and asymptotic variances must be equal for the correct $\mPS^2$.
In accordance with our prediction above, we see that the two variances coincide at the special choice $\mPS^2 = \mGaussNL^2$, indicating that a self-consistent study of the vacuum should include a modification of the initial conditions.
This point also appears to coincide with the minimum of the asymptotic variance.
A possible explanation for this is that rescattering effects remain subdominant so that each Fourier mode effective acts as an independent harmonic oscillator with frequency set by the renormalized effective mass.
In this case, for each Fourier mode the pseudovacuum initial conditions with smallest time average field variance is the one where $\mPS^2+k^2$ matches the oscillation frequency.

The coarse-grained view provided by the field variance $\fldVar$ indicates that the bare lattice vacuum lacks time-translation invariance when accounting for the effects of interactions between realized fluctuations.
We now take a more fine-grained look at the field fluctuations and study the evolution of the individual Fourier modes.
In~\secref{sec:mass-renorm} we showed that the potential force acting on the \emph{homogeneous} mode of the field receives a correction that can be interpreted as a shift in the effective mass.
Although our preceding result is not strictly applicable, we expect that inhomogeneous modes will experience a corresponding shift in their effective mass.
In fact, for symmetric minima we have shown that truncating the dynamical equation for $\delta\phi$ at cubic order in the amplitude (\ie the leading non-linear correction), produces a shift in the frequency of the modes whose $\fldVar$ dependence matches that of the one-loop approximation~\eqref{eqn:m2eff-one-loop-it} above.

\begin{figure}
  \includegraphics[width=4.5in]{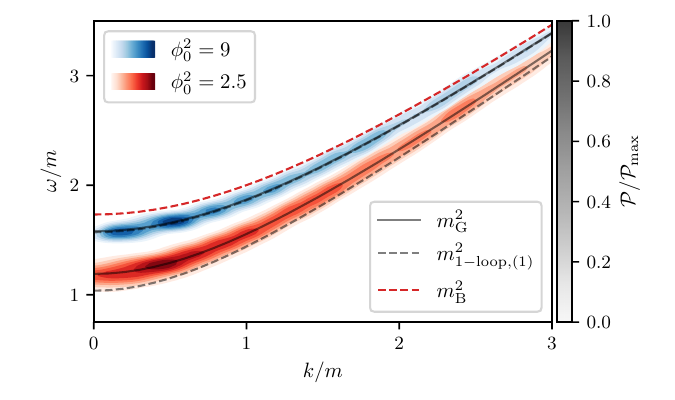}
  \caption{A demonstration of the numerical dispersion relationship using the power spectral density $\mathcal{P}$.  Here we have self-consistently realized the fluctuations with $\mPS^2 = m_{\rm eff}^2$.  Although not shown explicitly, we confirmed that very similar results are obtained by realizing bare vacuum fluctuations with $\mPS^2 = \vBare''(\phiFV)$, at least for the range of parameters shown in this plot. The solid grey lines are the analytic predictions for $\meff^2$ using the Gaussian approximation, while the dashed lines are using the one-loop bare mass approximation.  We show results for $\fldBEC^2 = 9$ (\emph{blue}) and $\fldBEC^2 = 2.5$ (\emph{red}).  In the former case, the one-loop and Gaussian resummed approximations match, and both accurately describe the results of the simulation.  However, for the latter case the one-loop approximation deviates noticably from the simulation results, while the Gaussian resummation continues to accurately capture the shifted frequency. For reference, the bare effective mass prediction is shown as a red dashed line.  All simulations used a lattice size $\mBEC L = 64$, $N=1024$ lattice sites, spectral cutoff $\nCut = 511$, and time step $dt = \frac{dx}{8}$.  The potential parameter was $\lamBEC = 2$. The averaging was performed over ensembles of $100$ simulations.  We remove the (rare) simulations that nucleate bubbles through a cut on $\ave{\cos\phi}$.  To reduce sampling artifacts, we interpolated the data points onto curves of constant $y \equiv \omega - \sqrt{\mGaussNL^2+k^2}$ and then applied a Gaussian filter along the constant $y$ lines.}
  \label{fig:dispersion-plot}
\end{figure}

We illustrate the mass shift in \figref{fig:dispersion-plot}, where we show the power spectral density
\begin{equation}
  \mathcal{P}(\omega,k) \propto \aveE{\abs{\delta\tilde{\phi}(\omega,k)}^2} \, ,
\end{equation}
with $\delta\tilde{\phi} \propto \int \ud t\ud x\, e^{ikx-i\omega t}\phi(k,t)$.  In these spectra, the presence of small amplitude oscillatory modes with a well-defined dispersion relationship $\omega(k)$ manifests as a sharp feature that traces out the line $(k,\omega(k))$.
For concreteness, we show the results from simulations using the corrected $\mPS^2 = \meff^2$ determined as described above.\footnote{We also ran simulations for initial conditions specified by the bare vacuum mass, $\mPS^2 = \vBare''(\phiFV) = \lamBEC^2-1$, and we found similar results for the model and lattice parameters used here.  This correspondence breaks down if we take $\fldBEC$ sufficiently small.  A possible explanation is that $\sigma^2_{\infty}$ is relatively flat around $\meff^2$, so that for initial bare masses near the corrected mass the effective mass of the field is insensitive to the initial value of the field fluctuations.}

\Figref{fig:dispersion-plot} clearly demonstrates the presence of a relativistic dispersion relationship $\omega^2 = k^2 + \meff^2$ for the inhomogeneous field fluctuations.  Further, the value of $\meff^2$ agrees with our predictions in the preceding section, indicating a distortion of the effective mass of the field due to the presence of the fluctuations.
We also see that our nonlinear Gaussian resummation~\eqref{eqn:m2eff-gauss-nl} continues to be valid even in regimes where the one-loop approximation has failed.

\section{Renormalization and the Euclidean Decay Rate}\label{sec:euclidean-actions}
In the preceding sections we showed that the effective dynamics for the lattice mean field is modified from that encoded in the bare lattice potential.
This raises interesting questions about the relationship between the bare potential, $\vBare$, appearing in our lattice simulations and the tree level potential, $V_{\rm T}$, appearing in the bounce calculation.
Both the real-time and Euclidean approaches approximate the full quantum dynamics using an expansion in $\hbar$.~\cite{Braden:2018tky,Hertzberg:2019wgx,Coleman:1977py,Callan:1977pt}.  Therefore, to compare the two formalisms, we should truncate them at the same order in $\hbar$.  With this in mind, let us briefly review the nature of both expansions.

Recall that in the Euclidean formalism (see, for example, refs.~\cite{Coleman:1977py,Callan:1977pt,Weinberg:2012pjx}), the decay rate per unit length takes the form
\begin{equation}\label{eqn:euclidean-decay-rate}
  \frac{\Gamma}{L} \approx \mu^2\left(\frac{S_{\rm E}}{2\pi\hbar}\right)e^{-\frac{S_{\rm E}}{\hbar}} \times \bar{D} \, ,
\end{equation}
where we have temporarily restored $\hbar$.
Here $S_{\rm E}$ is the Euclidean bounce action, obtained by solving the bounce equation, and $\bar{D}$ is a (dimensionless) prefactor encoding information about fluctuations about the bounce.  We will refer to $\bar{D}$ as the fluctuation determinant below.
$\mu$ is some inverse length scale characterizing the bounce that has been factored to make $\bar{D}$ dimensionless.
When we set $\hbar=1$, $\mu$ becomes a mass scale and will be referred to as such below.
The leading order contribution in $\hbar$ appears nonperturbatively in the $e^{-\frac{S_{\rm E}}{\hbar}}$ factor.
Meanwhile, the first quantum correction is encoded in the fluctuation determinant, which is expected to have the form $\bar{D} \sim 1+\mathcal{O}(\hbar)$.
As usual in quantum field theory, this determinant is naively divergent and must be renormalized.

Now, compare this to the real-time semiclassical lattice calculations.  The classical evolution captures contributions of classical trajectories to the path integral, which appear as $e^{\frac{i}{\hbar}S_{\rm cl}}$.  Viewed as an $\hbar$ expansion, this appears at the same order as the leading order contribution in the Euclidean formalism.
Further, in our lattice simulations the initial fluctuations have variance $\mathcal{O}(\hbar)$, and thus contribute at the same order as the corrections due to fluctuation determinant.
Combined with the fact that $\bar{D}$ encodes information about fluctuations around a highly symmetric saddle point (the bounce), this suggests that the lattice effective potential force is capturing the renormalization effects present in the fluctuation determinant.

This comparison of expansion orders suggests that Euclidean calculations must include renormalization effects in order to make a fair comparison with the real-time results.\footnote{More precisely, the shifts in the bare parameters observed in the lattice simulations must be accounted for in the Euclidean calculations.}
However, the standard Euclidean expansion~\eqref{eqn:euclidean-decay-rate} only includes one-loop quantum corrections, which as demonstrated above fail to properly describe the effective lattice potential.
A reasonable conjecture is that we should identify the effective lattice potential with an ``effective tree-level Euclidean potential'' $V_{\rm E,eff}$, and perform our bounce calculations in $V_{\rm E,eff}$.  Since the lattice results automatically include a UV cutoff, this substitution may capture the effects of the divergent part of the fluctuation determinant in the usual Euclidean formalism.
We leave a detailed study of the applicability of these conjectures to future work, and here we simply explore a few potential consequences if they are true.

Examining the Euclidean decay rate~\eqref{eqn:euclidean-decay-rate}, we see that fluctuation corrections can impact the predicted Euclidean rate in three ways:
\begin{enumerate}
  \item  modifying the Euclidean action $S_{\rm E}$,
  \item modifying the characteristic mass scale $\mu$, or
  \item modifying the $\mathcal{O}(\hbar)$ correction.
\end{enumerate}
Roughly, we expect that $\mu^2 \sim \abs{V''_{\rm eff, max}}$ is set by the curvature of the potential at the local maximum, which receives corrections just as the vacuum masses do.  However, for technical reasons, this shift is much more difficult to compute than masses at the symmetric local minima.
Similarly, understanding the modification to the $\mathcal{O}(\hbar)$ piece requires a detailed study of the fluctuation determinant.
Since our purpose is to get a flavor of the types of corrections we can expect, we will focus only on the changes to $S_{\rm E}$.
A reasonable guess is that $\mu^2$'s behavior may be similar to $\mFV^2$ and $\mTV^2$.
However, for this model even simple combinations such as $\mTV^2-\mFV^2$ and $\mTV^2+\mFV^2$ have different $\fldBEC$ dependence, so that more detailed investigation is needed.

With this in mind, we now explore how the Euclidean bounce action changes if we replace the bare lattice potential with an (approximate) effective lattice potential.
In~\secref{sec:mass-renorm} we showed how the curvature of the effective lattice potential is modified at a symmetric local minimum of the potential.
A first principles derivation of the effective potential requires extending these techniques to locations on the bare potential that are neither minima nor posess even symmetry.
This significantly complicates the required analysis, and we leave a more detailed development of the theory to future work.
Here, we instead take an effective theory (EFT) approach and expand the effective potential in a set of basis functions that respect the symmetries of the bare lattice potential.
The unknown coefficients in this expansion must then be set by measurements.
In keeping with our EFT inspired approach, we compute $\mTV^2$ and $\mFV^2$ by solving for the Gaussian resummed masses at $\phi=\fldBEC\pi$ and $\phi = 0$, respectively.
These numerical ``measurements'' are then used to fix two of the coefficients in our effective potential expansions.
Our potential parametrizations are specified by the number of basis functions we retain.
Within a fixed parametrization we can explore how modifying the properties of the fluctuations translates into the Euclidean action.
Alternatively, we can test the sensitivity of our results to modeling choices by allowing some of the unfixed parameters to float, providing us with a ``theory error bar'' of sorts.
The interested reader can find more details of the potential modeling in \appref{app:potential-modelling}.
These parametrizations are not meant to be fully realistic, but rather to give a flavor of the impact that using the effective lattice potential in tree level bounce calculations can have on the inferred decay rate.

\begin{figure}
  \includegraphics[width=7.05in]{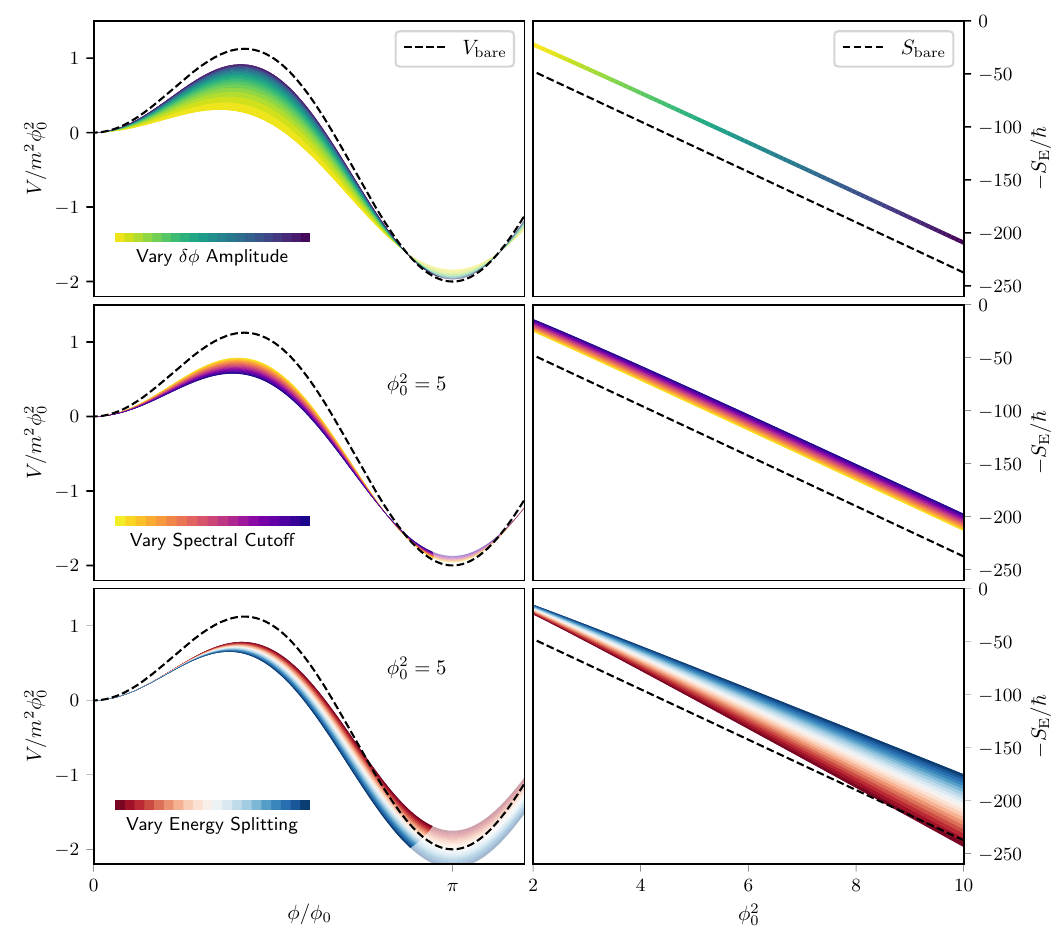}
  \caption{An illustration of the impact of potential deformations on the Euclidean action obtained in the bounce formalism.  In all cases, we use a bare lattice potential with $\lamBEC = 2$ and IR spectral cutoff $k_{\rm IR} = \frac{\pi}{64}$.  Unless otherwise specified, the fluctuations have a UV spectral cutoff $k_{\rm UV} = \frac{2\pi}{64}\times 511.5$.  In the left column, we illustrate the assumed potential deformations.  Meanwhile, the right column shows the corresponding response of the Euclidean action.  The first two rows show the impact of changing the fluctuation properties, assuming the two-term cosine expansion is valid, while the final row demonstrates some of the modeling uncertainty by considering a three-term cosine expansion. The Euclidean action is only directly influenced by the parts of the potential probed by the bounce solution.  We have indicated these regions with dark shading in the left plots.  The extension of the potentials beyond the region probed by the bounce are indicated with partially transparent shading. \emph{Top Row}: We vary $\fldBEC$, thus adjusting the fluctuation amplitude relative to the bare curvature scale of the false vacuum.  The effective potentials in the left panel are color coded by the value of $\fldBEC$, which match the corresponding Euclidean actions in the right panel. For reference, the bare lattice potential and corresponding Euclidean action are shown as a black dashed line.  \emph{Middle Row}: In the middle row we show the impact of adjusting the spectral cutoff of the fluctuations.  The left panel shows how the effective potential changes at a fixed value $\fldBEC^2 = 5$, again assuming the two term cosine expansion.  The color-coding increases in darkness as $\kUV$ is increased.   The response in the Euclidean for the same values of $\kUV$, but scanning across $\fldBEC$ values, is shown in the right panel.  \emph{Bottom Row}: Finally, the bottom row demonstrates one possible impact of potential modeling uncertainty by extending to the three-term cosine expansion.  For concreteness, we fix the new free parameter by tuning the dimensionless energy splitting $\frac{\Delta\rho}{\fldBEC^2\mBEC^2}$, at least in the range of $\fldBEC$ values shown in the right plot.  The color coding of the potentials signifies different choices of $\frac{\Delta\rho}{\fldBEC^2\mBEC^2}$ at fixed $\fldBEC$, with matching coloring of the Euclidean actions (where $\fldBEC$ is varied).}
  \label{fig:euclidean-actions}
\end{figure}

In the first two rows of~\figref{fig:euclidean-actions} we show the response of the Euclidean action to changes in the fluctuation statistics.
For illustration we use the two term truncation of the cosine series, allowing us to uniquely fix the potential via determination of $\mFV^2$ and $\mTV^2$.
In the absence of corrections and in one spatial dimension, the Euclidean action has the form
\begin{equation}
  S_{\rm bare} = \fldBEC^2  C(\lamBEC) \, ,
\end{equation}
where $C$ depends only on $\lamBEC$, and not $\fldBEC$ or $\mBEC$.
In the top row of the figure we fix the dynamic range of fluctuations, encoded in $\kIR$ and $\kUV$, while allowing $\fldBEC$ to vary.
This adjusts the field variance relative to the curvature scale of the potential minimum.
Despite generating relatively large corrections to the effective potential, we see that (in the range of $\fldBEC^2$ values shown in the plot) the main effect of the corrections is to shift $S_{\rm E}$ by a $\fldBEC^2$ independent constant.
In other words, in the leading approximation to $\ln\Gamma$ as a function of $\fldBEC^2$, the intercept with respect to $\fldBEC^2$ shifts while holding the slope fixed.
Of course, since the two curves must converge as $\fldBEC^2 \to \infty$, this cannot be a complete description, and we confirmed that for $\phi_0^2 \to \infty$, the two curves converge.

In the middle row, we additionally allow the UV cutoff on the fluctuations to vary.
The corresponding changes to $V_{\rm eff}$ at a fixed value of $\fldBEC^2$ while $\kUV^2$ is varied (in logarithmically spaced intervals) are illustrated in the left panel.
Somewhat surprisingly, we find the behaviour observed in the first row persists throughout the range of $\kUV$ values considered here, with the intercept of $-S_{\rm E}$ with respect to $\fldBEC^2$ increasing with $\kUV$.

A priori, it is rather remarkable that the seemingly complicated changes to the effective potential manifest in such a simple way in the Euclidean action.
However, we can provide a partial understanding applicable in the regime where the mass deformations are sufficiently small.
From our previous analysis, we know that $\Delta m^2 \approx -\alpha\fldBEC^{-2}$ for some constant $\alpha$ and sufficiently large $\fldBEC$.
For dimensional reasons, we expect $\Delta S \sim \Delta m^2 \fldBEC^2 \sim -\alpha$, resulting in a constant positive shift in $\Delta S$.
The detailed calculations provided in~\figref{fig:euclidean-actions} seem to indicate that similar scalings apply even when the individual mass corrections no longer scale as $\fldBEC^{-2}$, as shown in~\figref{fig:m2eff-prediction}.

In the investigations described above, we truncated our cosine expansions at second-order, leaving only two unknown coefficients.
Specifying two independent mass measurements uniquely determined the resulting effective potential, so that any uncertainties in this specific truncation arose solely from corrections to our computations of the effective masses.
However, since our effective potentials are not derived from first principles, there is significant modeling uncertainty.
Some obvious examples of this are our choice of truncation order and our choice to directly expand the effective potential (instead of another quantity such as $\fldVar$) in a cosine series.
To illustrate one possible effect of this modeling uncertainty, in the bottom row of~\figref{fig:euclidean-actions} we instead truncate the expansion at third order, introducing a single tunable parameter.
As outlined in~\appref{app:potential-modelling}, it is convenient to take this free parameter to be the vacuum energy splitting, as it is both physically transparent and straightforward to directly relate to the expansion coefficients.
Since the energy splitting can be expressed as
\begin{equation}
  \Delta\rho = \int_{\phiTV}^{\phiFV}F_{\rm eff}(\phi) \ud\phi\, ,
\end{equation}
it encodes (integrated) information about the shape of the potential force between the two minima.
In the bottow row of~\figref{fig:euclidean-actions}, we allow the dimensionless energy splitting $\frac{\Delta\rho}{\mBEC^2\fldBEC^2}$ to vary, resulting in the potential deformations shown in the left panel.
From the right panel, we see that unlike the previous cases, the primary effect of these variations is to change the slope of $S_{\rm E}$ with respect to $\fldBEC^2$.
Of course, such modification can only apply in a limited window of $\fldBEC$, since as $\fldBEC \to \infty$ at fixed $\kUV$ we must recover the bare potential, for which the vacuum energy splitting $\Delta\rho = 2\mBEC^2\fldBEC^2$ is known.

\section{Fluctuation Dependence of Real-Time Decay Rates}\label{sec:real-time-rates}
As shown above, the effective frequency of small amplitude oscillations around local potential minima is modified by interactions with the realized vacuum fluctuations.
The modification is well parametrized by an effective mass squared $\meff^2$ in a relativistic dispersion relationship
\begin{equation}
  \omega_k^2 = k^2 + \meff^2 \, .
\end{equation}
The effective mass squared is itself determined by the properties of the initial fluctuations.
In a similar fashion, we expect that the effective potential for the long-wavelength field modes will be modified, with a resulting change to the measured real-time decay rates.
To investigate this, we now show how the real-time decay rates respond to two natural modifications of the initial fluctuation state.  Analysis for a broader range of parameters will be presented in future work.

Before presenting the results, we briefly summarize our procedure to extract the decay rate.  We first determine the volume average $\left\langle\cos(\phi/\fldBEC)\right\rangle_{\rm V} \equiv c$ for each simulation.  We then define the decay time $t_{\rm decay}$ of a single simulation to be the first time that $c$ crosses the threshold value $c_{\rm cut} = 0.7$.  
From the collection of decay times $t_{\rm decay}^{\rm (i)}$ for a given ensemble, we compute the empirical survival probability $P_{\rm survive}$ as a function of time.  
Finally, we carry out a linear fit $\ln P_{\rm survive} = -\Gamma t + A$ to extract $\Gamma$.\footnote{For some choices of $\kCut$ and $\fldBEC^2$, the survival probabilities displayed two different exponentially decaying regimes.
In these cases, we have treated the earlier exponential as a transient and fit the later time exponential.
Specifically, if the survival fraction after one-light crossing time was less that $0.1$, we fit to the region $0.1 \leq P_{\rm survive} \leq 0.6$.
Otherwise, we use all of the decays with $P_{\rm survive} < 0.99$.
We verified that the extracted decay rate curves were insensitive to these particular choices of probability cutoffs (with the caveat that we fit only a single exponential region), as well as to the particular choice of threshold $c_{\rm cut}$.}
Our choice of spectral cutoffs was guided by the requirement that the nucleation times in each individual simulation were insensitive to the spatial resolution of the grid for a fixed choice of initial condition.

\begin{figure}
    \includegraphics[width=7.05826in]{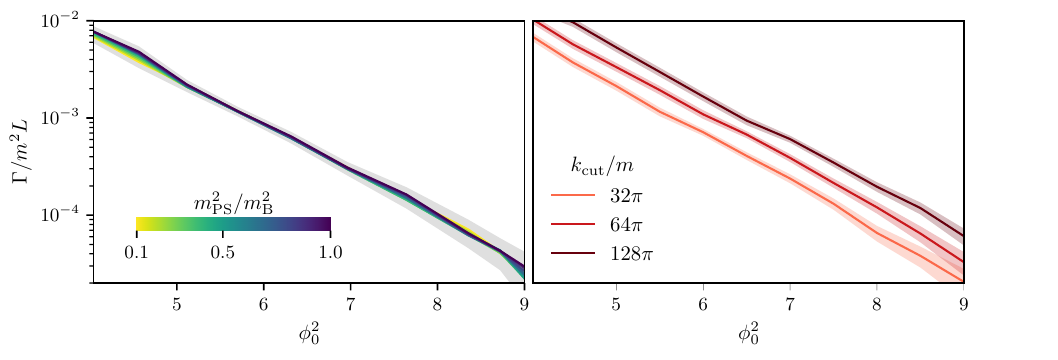}
    \caption{Decay rates (per unit volume) extracted from real-time lattice simulations for two deformations of the the initial conditions.  For illustration, and to compare with previous work, we have used a bare lattice potential with $\lambda=1.2$.  In all cases, we restricted our fits to times less than the light crossing time, although we verified that our results were insensitive to using decays within three light-crossing times.  \emph{Left}: We vary the pseudovacuum initial conditions by scanning the mass parameter $\mPS^2/\mBare^2$.  Since the corrections to $\mBare^2$ are negative for our fiducial potential, we vary $\mPS^2/\mBare^2$ in the interval $[0.1,1]$ in steps of size $\Delta\mPS^2/\mBare^2 = 0.1$.  We see that the extracted decay rates have almost no sensitivity to $\mPS^2$ for the parameters used here.  For these simulations, we used a spectral cutoff $\kCut = \frac{1536\pi}{25\sqrt{2}}$ and a box length $mL = 25\sqrt{2}$.  Each data point was extracted from a ensemble of $1000$ simulations.  \emph{Right:}  The measured decay rates as we vary the spectral cutoff $\kCut$, holding all other model and lattice parameters fixed.  For these simulations, we used a box size $mL = 32$ on pseudovacuum initial conditions with bare vacuum mass $\mPS^2 = m^2(\lambda^2-1)$.  Based on the left panel, we do not expect the results to depend on the choice of $\mPS^2$.  The change in the decay rate is approximated by a vertical shift in $\ln(\Gamma/m^2L)$ while holding the slope with respect to $\phi_0^2$ fixed.  This agrees with the qualitative behavior of the Euclidean rates in~\figref{fig:euclidean-actions}.  The decay rates were extracted using ensembles of $3000$ simulations.  The light shaded regions are an estimate of the statistical errors associated with the finite size of our ensembles.}
    \label{fig:real-time-decay-rates}
\end{figure}

Since the vacuum state of a free harmonic oscillator is dictated by its oscillation frequency, an accurate model of the Gaussian false vacuum should use pseudovacuum initial conditions with the self-consistently determined $\mPS^2 = \meff^2$.
However, existing analyses initialize the fluctuations based on linear perturbation theory with $\mPS^2 = \vBare''(\phiFV)$, thus effectively starting the simulations from an excited state.
Further, we demonstrated above that the field variance, and thus the effective mean field force, undergoes nontrivial dynamical evolution if the fluctuations are initialized away from the self-consistently determined vacuum.
It is therefore of interest to understand if the decay rate is sensitive to the choice of $\mPS^2$.
The left panel of~\figref{fig:real-time-decay-rates} explores this for a range pseudovacua initial conditions with $\mPS^2 \in [0.1\mBare^2,\mBare^2]$ and the particular choice $\lamBEC = 1.2$.
Remarkably, we see that the measured decay rate is independent of the initial $\mPS^2$, at least over the range of values considered here,
which indicates that the measured decay rates are robust to deformations of the vacuum state arising from modifications of the effective potential.
This is despite the fact that we expect the fluctuations to dynamically rearrange themselves, and we have effectively started the system in an excited state.
A possible explanation is that the asymptotic field variance is relatively flat over the range of mass deformations we simulated, as seen (for a different choice of $\lambda$) in~\figref{fig:variance-plot}.
As a result, the system may be rapidly equilibrating itself to a near-vacuum state that is not so different from the Gaussian vacuum.
A more definitive understanding requires further investigation.

Finally, we examine the dependence of the real-time decay rates on the spectral cutoff $\kCut$ used in the lattice simulations in the right panel of~\figref{fig:real-time-decay-rates}.
By increasing the cutoff, we are explicitly modeling more and more of the short-wavelength modes, rather than assuming they can be captured by simple modifications to the IR dynamics.
This type of modification of the initial conditions is of great interest, as it connects directly to the standard view of renormalization.
Analogous to the variation in the pseudovacuum initial conditions, we considered a range of values for $\kCut$ in the model with $\lamBEC = 1.2$.
Within the range of $\fldBEC$ values considered here, we see that impact of changing $\kCut$ is well modeled as a constant shift in the numerical value of $\ln\Gamma$ at fixed $\fldBEC^2$, while maintaining its slope as a function of $\fldBEC^2$.
This qualitative behaviour matches what we found for the decay rates predicted by the Euclidean formalism as the UV cutoff $\kUV$ was increased.
We leave a more detailed comparison between the $\kCut$ dependence in the real-time simulations and the $\kUV$ dependence of the Euclidean rates to future work.
For now we simply note that this can provide either a nontrivial way to differentiate the predictions of the two approaches, or an additional parameter to constrain our effective potential construction.

\section{Conclusions}\label{sec:conclusion}
We studied renormalization effects that result from realized vacuum fluctuations in nonlinear lattice simulations. Our focus is on relativistic scalar fields sitting at local potential minima.
First, we derived a quantitative prediction for the shift in the effective mass of the field and compared this prediction to measurements of oscillation frequencies in lattice simulations, finding excellent agreement.
Further, we demonstrated the variance of the field fluctuations was constant in time for this special choice of mass---a necessary condition for a vacuum state.
We then used our calculations of the mass modifications to partially fix the shape of the modified effective potential and studied the resulting changes to the false vacuum decay rates $\Gamma$ predicted by the Euclidean formalism, showing reasonable deformations can both shift the overall scale and slope of $\ln\Gamma$ with respect to $\phi_0^2$. 
Finally, we considered the sensitivity of decay rates measured in real-time simulations to variations of the effective mass ($\mPS^2$) and spectral cutoff ($\kCut$) defining the initial fluctuation spectra.
The measured decay rates were insensitive to the specific choice of $\mPS^2$, but had a similar dependence on $\kCut$ as the Euclidean rates.
Fully understanding the impact of these fluctuations on predictions of measurable quantities (such as decay rates) is important for comparing the results of real-time lattice simulations both with other theoretical approaches and with experiments.
In light of upcoming analog relativistic false vacuum decay experiments~\cite{Fialko:2014xba,Fialko:2016ggg,Braden:2017add,Billam:2018pvp,Braden:2019vsw,Billam:2020xna,Ng:2020pxk,Billam:2021qwt,Billam:2021nbc,Billam:2021psh}, these studies are particularly important to make optimal use of the experimental results.

Under the assumption of Gaussian fluctuations, we obtained the form of the effective potential force in terms of the field fluctuation variance.
From this, we derived the expected change in the effective mass of the field at the local minima of the potential.
We compared these analytic predictions to full nonlinear lattice simulations, and found excellent agreement across the range of parameters explored here.
In particular, the regime of validity of our Gaussian resummation approach was significantly larger than the one-loop approximation.
If we allow for increasingly large fluctuation amplitudes, we expect non-Gaussian corrections to become important, but we leave an exploration of this regime to future work.

Motivated by these observed modifications to the effective mass, we studied possible impacts on the decay rate predicted by the Euclidean formalism.
The periodicity and symmetry properties of the bare potential motivated an expansion of the effective potential in a cosine basis.
Truncating this expansion at second order allowed us to fix the potential coefficients using the true and false vacuum masses predicted by our formalism.
Under these assumptions, we found that the primary effect of the renormalization corrections was to shift the Euclidean action by a $\fldBEC$ independent, but $\kCut$ dependent constant.
Finally, we explored sensitivity to our specific modeling choice by including an additional term in the expansion and fixing it by allowing the vacuum energy splitting to vary.
We found that this modification adjusted the slope of the Euclidean action with respect to $\fldBEC^2$.
These findings motivate a more detailed study of the impacts renormalization corrections can have on the Euclidean action, which we leave to future work.

Finally, we explored the impact of modifying the fluctuation content on the decay rates extracted from lattice simulations.
In particular, we considered the sensitivity of the decay rate to two deformations of the initial conditions: (i) a variation in the initial pseudovacuum initial mass, and (ii) changes to the spectral cutoff $\kCut$.
We found the decay rate was insensitive to the initial mass used in the spectrum.
Meanwhile, increasing the spectral cutoff $\kCut$ led to an approximately $\fldBEC$ independent shift in the logarithmic decay rate.
This latter observation matches the qualitative behavior observed in the Euclidean decay rates under variations of $\kCut$.

This work suggests a number of future directions.
While we presented some simplified models for the renormalization effects in the effective potential, it would be interesting to perform a more systematic construction of the potential away from the minimum, including the effects of the non-Gaussian contributions neglected here.
This includes pushing the fluctuations into a regime where the Gaussian approximation begins to fail.
It is also of interest to understand how the fluctuations restructure themselves when starting from an excited pseudovacuum state.
Further, the potential considered here was special in the sense that the local minima were symmetric.
Extending the analysis to asymmetric minima introduces new effects, which will be presented in future work.
Finally, it is of interest to more rigorously explore the connection the effective lattice potential studied here, and the effective potential identified in Euclidean path integral calculations. 

\begin{acknowledgments}
The authors would like to thank Alexander Jenkins, Dalila P\^{i}rvu, Erick Weinberg, and Frank Wilczek for useful discussions.
The work of JB was partially supported by the Natural Sciences and Engineering Research Council  of Canada (NSERC).
JB was supported in part by the Simons Modern Inflationary Cosmology program.
This work was supported by the Science and Technology Facilities Council through the UKRI Quantum Technologies for Fundamental Physics Programme [grant numbers ST/T005904/1 and ST/T006900/1].
MCJ is supported by the National Science and Engineering Research Council through a Discovery grant. This research was supported in part by Perimeter Institute for Theoretical Physics. Research at Perimeter Institute is supported by the Government of Canada through the Department of Innovation, Science and Economic Development Canada and by the Province of Ontario through the Ministry of Research, Innovation and Science. This work was partly enabled by the UCL Cosmoparticle Initiative.  The data that support the findings of this study are available from the corresponding author, JB, under reasonable request.
\end{acknowledgments}

\appendix

\section{Modeling Effective Potential Corrections}\label{app:potential-modelling}
In this appendix we briefly present our models for the renormalization corrections arising from the presence of realized vacuum fluctuations on the lattice.  These models were used to study the sensitivity of the bounce action to RG corrections in~\secref{sec:euclidean-actions}.
A full nonperturbative analysis is beyond the scope of this paper, and the approximations presented here are meant to illustrate the effects that may arise in a more thorough analysis.

Before detailing our specific models, let's first consider some of the basic properties that the effective potential should posess, at least when considering the effect of ensemble averaged fluctuations.
The analog BEC potential considered here is periodic, and has even symmetry about both the false vacua (at $\phiFV = 2\pi i\fldBEC$ for $i\in\mathbb{Z}$) and true vacua (at $\phiTV = (2i+1)\pi\fldBEC$ for $i\in\mathbb{Z}$).
As a result, the effective potential should itself be periodic, with minima in at the same locations.
Similar considerations also extend to expansions of the field variance.

Given the requirements above, it is natural to expand the effective potential in a cosine basis
\begin{equation}\label{eqn:cosine-expansion}
  V_{\rm eff} = V_{\rm FV} + \sum_{n=1}^\infty c_n\left[\cos\left(n\frac{\phi}{\fldBEC}\right) - 1\right] \, ,
\end{equation}
where we have enforced $V_{\rm eff}(\phiFV) = V_{\rm FV}$ independent of the parameters $c_n$.
Alternatively, we can view this as an expansion of the effective force in a sine basis
\begin{equation}
  \forceEff = \sum_{n=1}^\infty -n\frac{c_n}{\fldBEC}\sin\left(n\frac{\phi}{\fldBEC}\right) \, .
\end{equation}
The parameters $c_n$ should be set either from detailed first principles calculations, or alternatively by calibrating the model to a collection of measurements.

\subsubsection{Calibration to Effective Mass Measurements}
A full first principles calculation of the $c_n$'s is beyond the scope of this paper.
However, we have an empirically verified theoretical understanding of the effective masses around symmetric potential minima.
Therefore, we take a more pedestrian approach and assume the squared effective masses at the false ($\mFV^2$) and true ($\mTV^2$) vacua are known, and use these to fix the unknown coefficients in our potential expansions.

For the case of the cosine expansion, the expansion coefficients $c_n$ must be calibrated so that $V''_{\rm eff}(0) = \mFV^2$ and $V''_{\rm eff}(\pi) = \mTV^2$.
At this level of approximation, we can then uniquely fix the potential by truncating after the $n=2$ term
\begin{equation}
  \frac{V_{\rm eff}}{\mBEC^2\fldBEC^2} = \frac{V_{\rm FV}}{\mBEC^2\fldBEC^2} + \frac{m_{\rm TV}^2-m_{\rm FV}^2}{2\mBEC^2}\left[\cos\left(\frac{\phi}{\fldBEC}\right)-1\right] - \frac{m_{\rm TV}^2+m_{\rm FV}^2}{8\mBEC^2}\left[\cos\left(2\frac{\phi}{\fldBEC}\right)-1\right] \, ,
\end{equation}
where the integration constant $V_{\rm FV}$ is fixed to the value of the potential at the false vacuum.
Alternatively, we could continue to incorporate additional information or uncertainties by including additional terms in the expansion.
For example, truncating at third order and assuming we have also measured the energy splitting between the false and true vacuum, we have
\begin{align}
  \frac{V_{\rm eff}}{\mBEC^2\fldBEC^2} &=
  \frac{V_{\rm FV}}{\mBEC^2\fldBEC^2} + \left(\frac{\mTV^2 - \mFV^2}{2\mBEC^2} - \frac{9}{16}\left[\frac{\mTV^2 - \mFV^2}{\mBEC^2} - \frac{\Delta\rho}{\mBEC^2\fldBEC^2}\right]\right)\left[\cos\left(\frac{\phi}{\fldBEC}\right)-1\right] \\
  &- \frac{\mTV^2 + \mFV^2}{8\mBEC^2}\left[\cos\left(2\frac{\phi}{\fldBEC}\right)-1 \right]
  + \frac{1}{16}\left[\frac{\mTV^2 - \mFV^2}{\mBEC^2} -\frac{\Delta\rho}{\mBEC^2\fldBEC^2}\right]\left[\cos\left(3\frac{\phi}{\fldBEC}\right)-1\right] \, ,
\end{align}
where
\begin{equation}
  \Delta\rho \equiv V_{\rm FV} - V_{\rm TV} = V_{\rm eff}(0) - V_{\rm eff}(\pi) \, .
\end{equation}
In the absence of corrections (\ie the bare limit), we have $(m_{\rm TV}^2 - m_{\rm FV}^2)\fldBEC^2 = 2\mBEC^2\fldBEC^2 = \Delta\rho$.
Making this substitution reduces the $n=3$ expansion to the $n=2$ expansion.

\bibliography{rg-effects-bec}
  
\end{document}